\documentclass{article}

\usepackage{epsfig} 
\usepackage{amsfonts}
\usepackage{graphicx}
\usepackage{amsmath}
\usepackage{mathtools}
\usepackage{cite}

 \usepackage{amssymb}

\date{\today}

\newcommand{\insertplot}[5]{\begin{figure}
 \hfill\hbox to 0.05in{\vbox to #5in{\vfill
 \inputplot{#1}{#4}{#5}}\hfill}
 \hfill\vspace{-.1in}
 \caption{#2}\label{#3}
 \end{figure}}
 \newcommand{\inputplot}[3]{
 \special{ps: plotfile #1}
}
\newcounter{fig}   

\numberwithin{equation}{section}

\newcommand{\ee}{\end{equation}}
\newcommand{\eea}{\end{eqnarray}}
\newcommand{\be}{\begin{equation}}
\newcommand{\bea}{\begin{eqnarray}}

\begin{document}

\title{\Large{ \bf 
Radiation from a $D$-dimensional collision of shock waves: \\
proof of first order formula and angular factorisation at all orders
}}

\author{
{\large Fl\'avio S. Coelho}\footnote{flavio@physics.org},  
{\large Carlos Herdeiro}\footnote{herdeiro@ua.pt}
and
{\large Marco O. P. Sampaio}\footnote{msampaio@ua.pt}
\\ 
\\
{\small Departamento de F\'\i sica da Universidade de Aveiro and I3N} \\ 
{\small   Campus de Santiago, 3810-183 Aveiro, Portugal}
}
\date{October 2014}
\maketitle

\begin{abstract} 
In two previous papers \cite{Herdeiro:2011ck,Coelho:2012sya} we have computed the inelasticity $\epsilon$ in  a head-on collision of two $D$-dimensional Aichelburg-Sexl shock waves, using perturbation theory to calculate the geometry in the future light-cone of the collision. The first order result, obtained as an accurate numerical fit, yielded the remarkably simple formula $\epsilon_{\rm 1st \ order}=1/2-1/D$. Here we show, \textit{analytically}, that this result is \textit{exact} in first order perturbation theory. Moreover, we clarify the relation between perturbation theory and an angular series of the inelasticity's angular power around the symmetry axis of the collision ($\theta=0,\pi$). To establish these results, firstly, we show that at null infinity the angular dependence factorises order by order in perturbation theory, as a result of a hidden symmetry. Secondly, we show that a consistent truncation of the angular series in powers of $\sin^{2}\theta$ at some order $\mathcal{O}(n)$ requires knowledge of the metric perturbations up to $\mathcal{O}(n + 1)$. In particular, this justifies the isotropy assumption used in first-order perturbation theory. We then compute, analytically, all terms that contribute to the inelasticity and depend linearly on the initial conditions (\textit{surface terms}), including second order contributions.
\end{abstract}

\newpage
 
\tableofcontents

\newpage

\section{Introduction}
The collision of two point-like particles at trans-Planckian energies should be well described by General Relativity~\cite{'tHooft:1987rb}. This problem is mostly of theoretical interest, to investigate the phenomenon of black hole (BH) creation in ultra-high energy collisions. Since the end of the XXth century, however, this problem also gained potential phenomenological relevance, with the advent of TeV gravity scenarios, based on higher-dimensional brane-world models~\cite{ArkaniHamed:1998rs,Antoniadis:1998ig,Randall:1999ee,Randall:1999vf}. These proposals led to suggestions of BH formation~\cite{Argyres:1998qn,Banks:1999gd}, both in realistic particle accelerators, namely at the LHC~\cite{Giddings:2001bu,Dimopoulos:2001hw}, and in ultra-high energy cosmic ray collisions~\cite{Feng:2001ib,Anchordoqui:2001ei,Emparan:2001kf}. Even though, so far, no signal of strong gravity has been found in the LHC data~\cite{Chatrchyan:2012taa,Aad:2012ic,Chatrchyan:2013xva,Aad:2013gma}, it remains of interest to obtain the best possible phenomenological description of such scenarios, as to produce reliable bounds on the fundamental Planck scale (see \textit{e.g.}, the discussion in~\cite{Cardoso:2012qm}).

A variety of analytical and numerical techniques have been used to model the collision of two particles/compact objects  at very high energies within $D$-dimensional General Relativity (see~\cite{Cardoso:2014uka} for a review). A most used (semi-)analytical technique is to take the two particles traveling precisely at the speed of light and model them as colliding Aichelburg-Sexl shock waves~\cite{Aichelburg:1970dh}. Then, \textit{bounds} for both the gravitational radiation emitted in the collision, as well as for the production cross section for BH formation, can be obtained by studying the formation of trapped surfaces in the collision of the two shock waves. This method, originally due to Penrose in four spacetime dimensions, has been extended to $D$ dimensions and generalized in different directions over the last decade~\cite{Giddings:2001bu,Yoshino:2002br,Yoshino:2002tx,Yoshino:2005hi,AlvarezGaume:2008fx,Albacete:2009ji}. We remark that BH formation does not occur in the case of other colliding impulsive shock waves, as in the classical example of Khan and Penrose~\cite{Khan:1971vh,Nutku:1977wp}.

A more ambitious program is to compute the metric in the future light-cone of the collision of the two Aichelburg-Sexl shock waves, from which \textit{estimates}, rather than bounds, can be obtained for the aforementioned observables. A perturbative framework for doing so, in $D=4$ and for a head-on collision, was constructed by D'Eath and Payne in a remarkable series of papers~\cite{DEath:1992hb,DEath:1992hd,DEath:1992qu}, building on earlier work by D'Eath~\cite{DEath:1990de}. The outcome of this program was the computation of the fraction of the total energy radiated away as gravitational waves, \textit{i.e.} \textit{the inelasticity}, $\epsilon$, in up to second order perturbation theory. Two encouraging observations for the validity of this approach are: (i) to this order the pre-collision exact solution of the Einstein equations is fully taken into account in the initial data; (ii) the result obtained by D'Eath and Payne, $\epsilon_{\rm 2nd \ order}=0.164$, agrees well with recent results from numerical relativity simulations colliding BHs~\cite{Sperhake:2008ga}, as well as other compact objects~\cite{East:2012mb,Rezzolla:2012nr}, at near the speed of light.

In a series of papers~\cite{Herdeiro:2011ck,Coelho:2012sya,Coelho:2012sy}, we have studied the head-on collision of two $D$-dimensional Aichelburg-Sexl shock waves, with the aim of computing the inelasticity, using an extension of the method of D'Eath and Payne to $D$ dimensions. The second order perturbation theory result still remains a technical challenge, and so far the main result of our study was to obtain a remarkably simple formula for the inelasticity, in \textit{first order perturbation theory}:
\begin{equation}
\epsilon_{\rm 1st \ order}=\frac{1}{2}-\frac{1}{D} \ .
\label{miracle}
\end{equation}
This formula was obtained as an accurate numerical fit~\cite{Coelho:2012sya}, since the integrals required to obtain the inelasticity could not be solved analytically. As such an obvious question arose: 
\begin{center}
{\bf Q1:} {\em is the result \eqref{miracle} \textit{exact} in first order perturbation theory?}
\end{center}
The first main goal of this paper is to answer, affirmatively, this question; we shall derive \eqref{miracle} by purely analytic means. Such analytic derivation is made possible through an asymptotic analysis and the use of Fourier space. In fact this procedure allows all surface term contributions to the inelasticity  to be computed analytically.

A central issue in the method of D'Eath and Payne is the relation between the perturbative expansion for the Einstein equations and a power series expansion for the inelasticity's angular power around the symmetry axis. In first order perturbation theory, for instance, following~\cite{DEath:1992hb}, we have used in~\cite{Herdeiro:2011ck,Coelho:2012sya} an isotropy assumption for the emission of gravitational radiation. Although some partial justifications for this assumption could be put forward, for instance from Smarr's `zero frequency limit'~\cite{Smarr:1977fy}, a question remained: 
\begin{center}
{\bf Q2:} {\em 
is the isotropy assumption well justified, for first order perturbation theory?}
\end{center}
The second main goal of this paper is to answer, again affirmatively, this question. Its answer is actually anchored to a central result that we shall prove and which clarifies the nature of the perturbative expansion in this method: 
\begin{center}
{\em for an asymptotic ``observer''Ä at null infinity, the angular dependence of the gravitational radiation signal factorizes order by order in perturbation theory, as a result of a hidden symmetry}.  
\end{center}
As a consequence of the analytic expression of this result, \textit{cf.} \eqref{Eq:angular_factorization2} and \eqref{Eq:angular_factorization},
we establish a correspondence between the order of the perturbative expansion and the order of the angular expansion of the metric close to the axis; in particular this allows us to answer Q2.

The structure of the paper is the following. We start by reviewing the perturbative setup and some of its general properties in $D$ dimensions. Thus, in Section~\ref{setup} we write down the formal solution to the perturbative series, decomposed in a suitable basis of functions, and in Section~\ref{formula} we introduce the (spectrally decomposed) inelasticity. In Section~\ref{CL} we extend the CL symmetry found in~\cite{DEath:1992hd} to $D$ dimensions, which allows for a partial separation of variables and reduction to two dimensions.  In Section~\ref{a_waveform} we prove the central result of the paper, quoted above. In Section~\ref{surface} we compute all surface terms, analytically, which exist up to second order in perturbation theory, at null infinity. In Section~\ref{conclusions} we present some final remarks. Many of the details of the calculations are left to the appendices.

\section{General setup and formal solution}
\label{setup}
In previous papers~\cite{Herdeiro:2011ck,Coelho:2012sya,Coelho:2012sy}, we have shown that the metric in the future light-cone of the head-on collision of two $D$-dimensional Aichelburg-Sexl shock waves can be constructed perturbatively with the ansatz
\begin{equation}
g_{\mu\nu}=\eta_{\mu\nu}+\sum_{k=1}^\infty h_{\mu\nu}^{(k)}\,, \qquad \eta_{\mu\nu}dx^\mu dx^\nu=-2dudv+\delta_{ij}dx^idx^j \ .\label{eq:pertexpansion}
\end{equation}
Here and throughout, the superscript index $(k)$ will be used to denote the order of the spacetime perturbation theory used to compute the given object. In particular, observe that for objects which are not linear in the metric perturbations, this does not correspond to the perturbation theory order of the object itself.

The ansatz \eqref{eq:pertexpansion} is subject to boundary data on a null hypersurface of the background Minkowski space, $u=0$, which contains all the information about the pre-collision geometry and is exact at second order (\textit{cf.} (3.2)--(3.6) in~\cite{Herdeiro:2011ck}). This construction is made in \textit{Brinkmann coordinates}, adapted to one of the shocks. In particular, this means that perturbation theory is valid around one particular end of the collision axis -- the negative $z$ axis, corresponding to $\theta=\pi$ in the discussion below and also in~\cite{Herdeiro:2011ck}, \textit{cf.} Fig. 1 therein (see also~\cite{Sampaio:2013faa} for an alternative argument for the validity of perturbation theory). Thus this perturbative expansion breaks the manifest $\mathbb{Z}_2$ symmetry of the problem (reflexions around $\theta=\pi/2$). This breakdown will manifest itself in the discussion below.

The vacuum Einstein field equations for the ansatz \eqref{eq:pertexpansion}  are written (in de Donder gauge) as a tower of wave equations with sources (\textit{cf.} (2.5) in~\cite{Coelho:2012sy}) 
\begin{equation}\label{eq:Feq}
\Box h^{(k)}_{\mu\nu}=T^{(k-1)}_{\mu\nu}\left[h_{\alpha\beta}^{(j<k)}\right] \, ,
\end{equation}
to be solved in the future light-cone of the collision. The right hand side source is generated, at each order, by the lower order perturbations which are assumed to have been computed already. This provides an iterative method to solve Einstein's equations order by order.  The general solution of~(\ref{eq:Feq}), at each order, using the Green's function method (see Theorem~6.3.1 of~\cite{Friedlander:112411}), is
\begin{equation}\label{GeneralInt}
h^{(k)}_{\mu\nu}(y)=F.P.\int_{u'>0}d^{D}y'\, G(y,y')\left[T^{(k-1)}_{\mu\nu}(y')+2\delta(u')\partial_{v'}h^{(k)}_{\mu\nu}(y')\right]\; .
\end{equation}
Here $F.P.$ denotes the finite part of the integral, $y=\left\{u,v,x^i\right\}$ and we have used the fact that the source only has support in $u>0$. We call the first (second) type of terms in \eqref{GeneralInt}, \textit{volume} (\textit{surface}) terms. Volume terms exist in second and higher order perturbation theory. Surface terms exist only in first and second order perturbation theory.

In~\cite{Coelho:2012sy}  we have made use of a basis of tensorial irreducible representations  of the axial, $SO(D-2)$, isometry, constructed from $x^i$ and $\delta_{ij}$. A convenient choice is ($\rho^2\equiv x_ix^i$)
\begin{equation}\label{eq:basis_tensors}
\Gamma_i\equiv \frac{x_i}{\rho} \; \;, \; \; \; \; \delta_{ij} \; \;, \; \; \; \; \Delta_{ij}\equiv \delta_{ij}-(D-2)\Gamma_i\Gamma_j \;,
\end{equation} 
where the last tensor is traceless. Then the metric perturbations, in de Donder coordinates, are decomposed into seven functions of $(u,v,\rho)$, here denoted $A,B,C,E,F,G,H$, as follows:
\begin{eqnarray}
&h_{uu}\equiv A=A^{(1)}+A^{(2)}+\ldots \qquad  &h_{ui}\equiv B \,\Gamma_i =(B^{(1)}+B^{(2)}+\ldots)\Gamma_i  \nonumber\\
&h_{uv}\equiv C=C^{(1)}+C^{(2)}+\ldots \qquad   &h_{vi}\equiv F \,\Gamma_i =(F^{(1)}+F^{(2)}+\ldots)\Gamma_i  \label{app:gen_perts} \hspace{-5mm}\\
&h_{vv}\equiv G=G^{(1)}+G^{(2)}+\ldots  \qquad &h_{ij}\equiv E \,\Delta_{ij}+H\, \delta_{ij} = (E^{(1)}+\ldots) \Delta_{ij}+(H^{(1)}+\ldots) \delta_{ij} \; .\nonumber 
\end{eqnarray}
The terms $A^{(k)},B^{(k)},C^{(k)},E^{(k)},F^{(k)},G^{(k)},H^{(k)}$ denote the $k^{th}$ order contribution, in perturbation theory, to the given function.

A similar decomposition can be applied to the sources $T^{(k-1)}_{\mu\nu}$. In particular, for the radiative components,
\begin{eqnarray}
T^{(k-1)}_{ij}&=&T_H^{(k-1)}(u,v,\rho)\delta_{ij}+T^{(k-1)}_E(u,v,\rho)\Delta_{ij}\,.
\end{eqnarray}
In what follows we denote a generic metric function projection (such as any of the functions in~\eqref{app:gen_perts}) by $F(u,v,\rho)$. Similarly, its associated source, \textit{i.e.} the corresponding projection of $T^{(k-1)}_{\mu\nu}$ with the same basis tensor, is denoted by $T(u,v,\rho)$. Then, after contracting~\eqref{GeneralInt} with the corresponding basis tensor, one obtains
\begin{equation}
F^{(k)}(u,v,\rho)=F.P.\int_{u'>0} d^{D}y'\, G(y,y')\Lambda_m\left(\frac{x\cdot x'}{\rho\rho'}\right)\left[T^{(k-1)}(u',v',\rho')+2\delta(u')\partial_{v'}F^{(k)}(0,v',\rho')\right]\,,\label{general_F}
\end{equation}
where the projection scalars can take three possible forms,
\begin{equation}
\Lambda_m(z)\equiv \left\{1,z,(D-3)^{-1}\left((D-2)z^2-1\right)\right\}\,,
\label{lm}
\end{equation}
respectively for $m=\{0,1,2\}$. Here and below $m$ is the rank of the basis tensor multiplying $F$ in the contribution to the metric perturbation (and similarly for $T$). Thus, $m=0$ for $A,C,G,H$, $m=1$ for $B,F$ and $m=2$ for $E$, as can be observed in~\eqref{app:gen_perts}.

In the following we denote by $S$ the sum of both volume and surface sources:
\begin{equation}
S^{(k)}(u,v,\rho)\equiv T^{(k-1)}(u,v,\rho)+2\delta(u)\partial_{v}F^{(k)}(0,v,\rho) \ .
\end{equation}

\subsection{The inelasticity from a Fourier representation}
\label{formula}
In~\cite{Coelho:2012sy}, we have derived a formula for the \textit{inelasticity} of the collision, \textit{i.e.} the percentage of the initial centre-of-mass energy that is radiated away as gravitational waves, using a $D$-dimensional generalisation of the Bondi mass loss formula~\cite{Tanabe:2011es,Tanabe:2012fg}. The Bondi coordinates $(r,\tau,\theta)$ used in the derivation relate to $(u,v,\rho)$ through\footnote{Strictly speaking, the metric in de Donder gauge, when written in $(r,\tau,\theta)$ coordinates, is only in Bondi gauge at null infinity, \textit{i.e.} in the limit $r\rightarrow\infty$.}
\begin{equation}
\sqrt{2}u=\tau+r(1-\cos\theta)\,,\qquad \sqrt{2}v=\tau+r(1+\cos\theta)\,,\qquad \rho=r\sin\theta\,,
\label{ct1}
\end{equation}
and the inelasticity, denoted by $\epsilon$, was shown to be (\textit{cf.} (2.21) in~\cite{Coelho:2012sy})
\begin{equation}
\epsilon=\int_{-1}^{1}\frac{d\cos\theta}{2}\,\epsilon(\theta)\equiv\int_{-1}^{1}\frac{d\cos\theta}{2}\int d\tau\, \dot{\mathcal{E}}(\tau,\theta)^2\,.\label{epsilon}
\end{equation}
$\epsilon(\theta)$ is the \textit{inelasticity's angular distribution}; the inelasticity's angular power $\dot{\mathcal{E}}(\tau,\theta)$ shall be called \textit{wave form} or, following the traditional nomenclature in $D=4$, the \textit{news function}. It reads
\begin{eqnarray}
\dot{\mathcal{E}}(\tau,\theta)&=&\lim_{r\rightarrow\infty}\frac{d}{d\tau}\mathcal{E}(r,\tau,\theta)\,, \label{awf}\\
\mathcal{E}(r,\tau,\theta)&=&\sqrt{\frac{(D-2)(D-3)}{4}}\,r\rho^{\frac{D-4}{2}}\left(E(u,v,\rho)+H(u,v,\rho)\right)\,,\label{E}
\end{eqnarray}
in terms of the functions introduced in \eqref{app:gen_perts}.

In Section \ref{a_waveform} we will find it convenient to work in Fourier space. Defining the Fourier transform of the wave form
\begin{equation}
\hat{\dot{\mathcal{E}}}(\omega,\theta)=\int d\tau\, \dot{\mathcal{E}}(\tau,\theta)\,e^{-i\omega\tau}\,,
\end{equation}
the angular distribution $\epsilon(\theta)$ can be expressed as a spectral integral using Plancherel's theorem
\begin{equation}
\epsilon(\theta)=\frac{1}{2\pi}\int d\omega\, |\hat{\dot{\mathcal{E}}}(\omega,\theta)|^2=\frac{1}{\pi}\int_0^{\infty} d\omega\, \hat{\dot{\mathcal{E}}}(\omega,\theta)\hat{\dot{\mathcal{E}}}(-\omega,\theta)\,.
\end{equation}
The last equality follows from the reality of $\dot{\mathcal{E}}(\tau,\theta)$.

In practice, since we are working within a perturbative framework, \eqref{E} will be computed at each order from the corresponding $E$ and $H$ at that order,
\begin{eqnarray}
\mathcal{E}^{(k)}(r,\tau,\theta)&=&\sqrt{\frac{(D-2)(D-3)}{4}}\,r\rho^{\frac{D-4}{2}}\left(E^{(k)}(u,v,\rho)+H^{(k)}(u,v,\rho)\right)\,.\label{E2}
\end{eqnarray}
The corresponding perturbative expansion for $\epsilon(\theta)$ is slightly different since, from \eqref{epsilon}, $\epsilon(\theta)$ is quadratic in the wave form. Introducing the perturbative expansion for the wave forms, we have
\begin{equation}
\epsilon(\theta)=\int d\tau\, \dot{\mathcal{E}}(\tau,\theta)^2=\int d\tau\, \sum_{k=1}^\infty \dot{\mathcal{E}}^{(k)}(\tau,\theta)\sum_{k'=1}^\infty \dot{\mathcal{E}}^{(k')}(\tau,\theta) \,,\label{epsilon2}
\end{equation}
so we see that the perturbative expansion for $\epsilon(\theta)$ takes the form
\begin{equation}
\epsilon(\theta)=\sum_{N=1}^\infty\epsilon^{(N)}(\theta)\,,
\label{eseries}
\end{equation}
where
\begin{equation}
\epsilon^{(N)}(\theta)=\sum_{k=1}^{N}\int d\tau\, \dot{\mathcal{E}}^{(k)}(\tau,\theta)\dot{\mathcal{E}}^{(N+1-k)}(\tau,\theta)\,.
\label{seriesi}
\end{equation}
Thus, knowledge of the metric perturbations to $\mathcal{O}(k)$ in spacetime perturbation theory allows us to consistently determine terms for the inelasticity's angular distribution (and hence to the inelasticity) up to $\mathcal{O}(k)$ in the series \eqref{eseries}. For instance, in first order perturbation theory we determined $\dot{\mathcal{E}}^{(1)}$ and thus $\epsilon^{(1)}(\theta)$; but $\epsilon^{(2)}(\theta)$ requires also knowledge of $\dot{\mathcal{E}}^{(2)}$.

\section{The CL symmetry}
\label{CL}

The head-on collision of two Aichelburg-Sexl shock waves contains, besides the $SO(D-2)$ isometry, a very useful extra symmetry, first pointed out in $D=4$ by D'Eath and Payne~\cite{DEath:1992hd}, which we now generalise to $D>4$. This allows for a partial separation of variables and a reduction to two dimensions.

\subsection{One shock wave}

Recall the metric of one Aichelburg-Sexl shock wave in \textit{Rosen coordinates} (\textit{cf.} (2.5) in~\cite{Herdeiro:2011ck}),
\begin{equation}
ds^2=-d\bar{u}d\bar{v}+\left(1+\kappa\frac{\bar{u}\theta(\bar{u})}{2}\Phi''(\bar{\rho})\right)^2d\bar{\rho}^2+\bar{\rho}^2\left(1+\kappa\frac{\bar{u}\theta(\bar{u})}{2\bar{\rho}}\Phi'(\bar{\rho})\right)^2d\bar{\Omega}^2_{D-3}\,,
\end{equation}
where the profile function, as defined in~\cite{Herdeiro:2011ck}, is
\begin{equation}
\Phi(\rho)=\left\{
\begin{array}{ll}
 -2\ln(\rho)\ , &  D=4\  \vspace{2mm}\\
\displaystyle{ \frac{2}{(D-4)\rho^{D-4}}}\ , & D>4\ \label{Phi}\,
\end{array} \right. \ .
\end{equation} 
It is well known that the only effect of a Lorentz transformation along the direction of the shock wave's motion is to rescale the energy parameter $\kappa$. Physically this is because such a transformation can only blue/red-shift the particle traveling at the speed of light. Since the energy is the only scale of the geometry, we can undo the Lorentz transformation, up to an overall conformal factor, by a further scaling of the coordinates. For concreteness, let $L$ be the Lorentz transformation $(\bar{u},\bar{v},\bar{x}^i)\xrightarrow{L}(e^{-\beta}\bar{u},e^\beta\bar{v},\bar{x}^i)$ and $C$ the conformal scaling $(\bar{u},\bar{v},\bar{x}^i)\xrightarrow{C}e^{-\frac{1}{D-3}\beta}(\bar{u},\bar{v},\bar{x}^i)$. Then, under $CL$,
\begin{eqnarray}
(\bar{u},\bar{v},\bar{x}^i)&\xrightarrow{CL}&(e^{-\frac{D-2}{D-3}\beta}\bar{u},e^{\frac{D-4}{D-3}\beta}\bar{v},e^{-\frac{1}{D-3}\beta}\bar{x}^i)\,,\\
g_{\mu\nu}(X)&\xrightarrow{CL}&e^{\frac{2}{D-3}\beta}g_{\mu\nu}(X')\; ,
\end{eqnarray}
so that the metric of a single shock wave simply scales by an overall constant factor under this transformation. Thus, this is a one parameter \textit{conformal isometry} of the background. One can then find spacetime coordinates adapted to the orbits of this conformal isometry. Such orbits naturally foliate the spacetime into $(D-1)$-dimensional leaves, which are left invariant by the action of the conformal symmetry. A suitable set of coordinates on such leaves are 
%
$\left\{p,q,\phi_i\right\}$, with
\begin{equation}
p\equiv \bar{v}\bar{\rho}^{D-4}\,,\qquad q\equiv \bar{u}\bar{\rho}^{2-D}\,, 
\label{leaves}
\end{equation}
and where $\phi_i$ are the angles on the transverse plane. The coordinate along the orbits of the conformal  symmetry, on the other hand, is simply $\bar{\rho}$, which transforms as  $\bar{\rho}\rightarrow\bar{\rho}'=e^{-\frac{1}{D-3}\beta}\bar{\rho}$.

Let us pause momentarily to remark that the presence of the $\bar{\rho}$ coordinate in (\ref{leaves}) is encouraging. Even though the problem has ``only'' axial symmetry, the conformal symmetry relates different $\bar{\rho}=$ constant cylinders, albeit at different $\bar{u},\bar{v}$. This symmetry plays a central role in the construction that follows in the next section, for which the first step is to understand the role of the CL symmetry in a spacetime with two colliding shock waves.

\subsection{Two colliding shock waves}

For the collision of two shock waves with energy parameters $\lambda$ and $\nu$, one can show, following~\cite{DEath:1992hd}, that under CL the perturbative expansion remains the same except for $\lambda\rightarrow \lambda e^{-2\beta}$. Thus, in the centre-of-mass frame, the expansion \eqref{eq:pertexpansion} becomes 
\begin{eqnarray}\label{eq:CLonMetric}
g_{\mu\nu}(X) \xrightarrow{CL}g_{\mu\nu}(X')=e^{\frac{2}{D-3}\beta}\left[\eta_{ \mu \nu}+\sum_{k=1}^\infty e^{-2k\beta}  h_{\mu\nu}^{(k)}(X')\right]\,.
\end{eqnarray}  
The metric functions $h_{\mu\nu}^{(k)}$ used in the perturbative expansion are the same, except that they are evaluated on the transformed coordinates on the right hand side, \textit{i.e.} before CL ($X$, left) and after CL ($X'$, right), respectively. 
On the other hand, a general coordinate transformation obeys $g_{\mu \nu}(X)=\frac{\partial x^{\mu'}}{\partial x^\mu}\frac{\partial x^{\nu'}}{\partial x^\nu}g_{\mu' \nu'}(X')$. In the present case ${\partial x^{\mu'}}/{\partial x^\mu}$ is the coordinate transformation operated by CL; thus, inserting in~\eqref{eq:CLonMetric}, we conclude that 
\begin{equation}\label{CL_transf}
h_{\mu\nu}^{(k)}(X')=e^{(2k+N_u-N_v)\beta}h_{\mu\nu}^{(k)}(X)\,,
\end{equation}
where $N_u$ and $N_v$ are the number of $u$-indices or $v$-indices\footnote{For example $h_{u'u'}=e^{(2k+2)(D-3)\beta}h_{uu}$, $h_{v'v'}=e^{(2k-2)(D-3)\beta}h_{vv}$ and $h_{i'j'}=e^{2k(D-3)\beta}h_{ij}$.}. Changing back to Brinkmann coordinates adapted to one of the shock waves\footnote{These are the coordinates in which the perturbative calculation in de Donder gauge is set up, \textit{cf.} Section~\ref{setup}  -- see~\cite{Herdeiro:2011ck} for the explicit coordinate transformation.}, the coordinates $p,q$ read
\begin{equation}
p=(\sqrt{2}v-\Phi(\rho))\rho^{D-4}\,,\qquad q=\sqrt{2}u\rho^{-(D-2)}\,.\label{def_pq}
\end{equation}
In these new \textit{CL-adapted} coordinates $\{\rho,p,q,\phi_i\}$ only $\rho$ transforms, i.e. $\rho\stackrel{CL}{\rightarrow} \rho'=e^{-\frac{1}{D-3}\beta}\rho$. This implies a separation of the $\rho$ variable in the form 
\begin{equation}\label{eq:separation_rho}
h_{\mu\nu}^{(k)}(p,q,\rho,\phi_i)=\frac{f^{(k)}_{\mu\nu}(p,q,\phi_i)}{\rho^{(D-3)(2k+N_u-N_v)}}\,,
\end{equation}
which can be checked by inserting~\eqref{eq:separation_rho} in~\eqref{CL_transf} on either side.

Next, and as in \eqref{app:gen_perts}, we can separate the (transverse plane) angular dependence by defining scalar functions analogous to $A^{(k)}(u,v,\rho)$, $E^{(k)}(u,v,\rho),\ldots$ for $f^{(k)}_{\mu\nu}$. 
We define such functions using the separation property~\eqref{eq:separation_rho}, which implies that, for each such function, generically denoted by $F$ as before,
\begin{equation}
F^{(k)}(u,v,\rho)= \dfrac{f^{(k)}(p,q)}{\rho^{(D-3)(2k+N_u-N_v)}}\,.\label{f_pq}
\end{equation}
Similarly, for the respective source $S^{(k)}(u,v,\rho)$, one shows that (using the scaling of the d'Alembertian operator under the symmetry in~\eqref{eq:Feq})
\begin{equation}
S^{(k)}(u,v,\rho)=\dfrac{\Sigma^{(k)}(p,q)}{\rho^{(D-3)(2k+N_u-N_v)+2}}\,.\label{def_Sigma}
\end{equation}
The bottom line is that, in these CL-adapted coordinates, the problem becomes two-dimensional order by order in perturbation theory, due to the factorization of the $\rho$ dependence. This is a computational advantage in the evaluation of both the surface and volume terms in~\eqref{general_F}, since one can factor out the $\rho'$ integrations and remove the $\rho$ dependence altogether from the integral. There is, however, another consequence of this symmetry which only reveals itself at null infinity and that will be discussed in the next section.

\section{Asymptotic metric functions and the news function}
\label{a_waveform}
Building on the previous section, we now derive two key results:
\begin{enumerate}
\item We prove that the angular dependence of all asymptotic metric functions factors out order by order in perturbation theory if we change to a new time coordinate, as a consequence of the CL symmetry at null infinity.
\item Then we obtain an expression for the inelasticity's angular distribution $\epsilon(\theta)$ as an angular series in powers of $\sin^2\theta$. Consequently, we show that there is a clear correspondence between the order of the terms in the angular series and the order of the perturbative expansion. This relation was previously claimed without rigorous proof~\cite{DEath:1992hb,DEath:1992hd,DEath:1992qu,Coelho:2012sy}.
\end{enumerate}

\subsection{Asymptotic factorization of the angular dependence}
\label{secfact}
We saw in Section \ref{formula} that the computation of the inelasticity is done over a sphere at null infinity, \textit{i.e.} it requires taking the limit $r\rightarrow\infty$ (\textit{cf.} \eqref{awf}). Using \eqref{ct1} and \eqref{def_pq}, the coordinates $p,q$ are related to $(r,\tau,\theta)$ through
\begin{equation}
p=(\tau+r(1+\cos\theta)-\Phi(r\sin\theta))(r\sin\theta)^{D-4}\,,\qquad q=(\tau+r(1-\cos\theta))(r\sin\theta)^{-(D-2)}\,.
\end{equation}

In the asymptotic limit, they behave as $q\rightarrow0$ and $p\rightarrow\infty$. It is reasonable to expect that the finite, non-trivial dependence of the wave form should be given in terms of combinations of $p,q$ that remain finite and non-trivial at null infinity. An appropriate choice is $\hat{p},\hat{q}$ given by
\begin{equation}
\hat{p}\equiv 
\left\{
\begin{array}{ll}
\displaystyle{ \frac{pq-1}{q}}+2\log q\ , &  D=4\  \vspace{2mm}\\
\displaystyle{ \frac{pq-1}{q^{\frac{1}{D-3}}}}\ , & D>4\ \label{tau_transform}\,
\end{array} \right. \ ,
\qquad \hat{q}\equiv q^{\frac{1}{D-3}}\,.
\end{equation}
This coordinate transformation, $(p,q)\rightarrow(\hat{p},\hat{q})$, has a constant Jacobian determinant, $D-3$, hence it is well defined everywhere. When $r\rightarrow\infty$,
\begin{equation}
\hat{p}\rightarrow2\bar{\tau}(\tau,\theta)+O\left(r^{-1}\right)\,,\qquad \hat{q}\rightarrow \frac{(1-\cos\theta)^{\frac{1}{D-3}}(\sin\theta)^{-\frac{D-2}{D-3}}}{r}+O\left(r^{-2}\right)\,,
\end{equation}
where we have defined a new time coordinate
\begin{eqnarray}
\bar{\tau}(\tau,\theta)=\left\{
\begin{array}{ll}
\displaystyle{ \frac{\tau}{1-\cos\theta}+\log\left(\frac{1-\cos\theta}{\sin\theta}\right)}\ , &  D=4\  \vspace{2mm}\\
\displaystyle{ \frac{\tau}{(1-\cos\theta)^{\frac{1}{D-3}}}(\sin\theta)^{-\frac{D-4}{D-3}}}\ , & D>4\ \label{tau_transform2}\,
\end{array} \right. \ .
\end{eqnarray}
Recall that we had already factored out the $\rho$ dependence, leaving only a function of $(p,q)$ to compute. But since $\bar\tau$ is the only surviving quantity in the asymptotic limit, we expect the wave form to become effectively one-dimensional at null infinity: besides the trivial dependence on $\theta$ coming from the known powers of $r$ and $\rho$, it must be a function of $\bar\tau(\tau,\theta)$ only. From the definitions \eqref{E2} and \eqref{f_pq},
\begin{equation}
\dot{\mathcal{E}}^{(k)}(r,\tau,\theta)=(\sin\theta)^{\frac{D-4}{2}-2k(D-3)}\times\frac{d}{d\tau}\left[r^{\frac{D-2}{2}-2k(D-3)}\hat{f}^{(k)}(\hat{p},\hat{q})\right]\,,
\end{equation}
where $\hat{f}^{(k)}$ contains the relevant contribution from $E^{(k)}$ and $H^{(k)}$, as a function of $(\hat{p},\hat{q})$. If the quantity inside brackets is to remain finite at null infinity, then necessarily
\begin{equation}
\hat{f}^{(k)}(\hat{p},\hat{q})\simeq\alpha^{(k)}(\hat{p})\, \hat{q}^{\frac{D-2}{2}-2k(D-3)}+\dots\,,
\end{equation}
where $\dots$ denotes higher powers of $\hat{q}$ and $\alpha^{(k)}(\hat{p})$ is some (yet) unknown function of $\hat{p}$. Thus, after taking the limit $r\rightarrow\infty$,
\begin{equation}
\dot{\mathcal{E}}^{(k)}(\tau,\theta)=\frac{1}{1-\cos\theta}\left(\frac{1+\cos\theta}{1-\cos\theta}\right)^{k-1+\frac{1}{4}\frac{D-4}{D-3}}\frac{d\bar{\tau}}{d\tau}(\theta)\frac{d}{d\bar{\tau}}\alpha^{(k)} (2\bar\tau)\,.
\end{equation}
The function $\alpha^{(k)}(2\bar\tau)$ can be given in terms of the wave form by evaluating the previous expression at $\theta=\tfrac{\pi}{2}$ and noting that $\bar\tau\left(\tau,\tfrac{\pi}{2}\right)=\tau$. Then,  we get
\begin{equation}
\dot{\mathcal{E}}^{(k)}(\tau,\theta)=\left(\frac{1}{1-\cos\theta}\right)^2\left(\frac{1+\cos\theta}{1-\cos\theta}\right)^{k-1-\frac{1}{4}\frac{D-4}{D-3}}\dot{\mathcal{E}}^{(k)}\left(\bar\tau(\tau,\theta),\frac{\pi}{2}\right)\, .\label{Eq:angular_factorization2}
\end{equation}
This proves the advertised factorization of the angular part of the wave form at null infinity, when the latter is written in terms of the new time coordinate $\bar{\tau}$.

Inserting \eqref{Eq:angular_factorization2} in the inelasticity power series \eqref{seriesi}, and changing the integration variable from $\tau$ to $\bar{\tau}$, we conclude that
\begin{equation}
\epsilon^{(N)}(\theta)=\left(\frac{1}{1-\cos\theta}\right)^3 \left(\frac{1+\cos\theta}{1-\cos\theta}\right)^{N-1}\epsilon^{(N)}\left(\frac{\pi}{2}\right)\, . \label{seriesi2}
\end{equation}
Thus it suffices to compute the inelasticity's angular distribution on the symmetry plane. The whole series reads
\begin{eqnarray}
\epsilon(\theta)&=&\sum_{N=1}^\infty \epsilon^{(N)}(\theta) =\left(\frac{1}{1-\cos\theta}\right)^3\sum_{N=1}^\infty\left(\frac{1+\cos\theta}{1-\cos\theta}\right)^{N-1}\epsilon^{(N)}\left(\frac{\pi}{2}\right)\,. 
\label{series2}
\end{eqnarray}

Two comments are in order. First, each term in this series is not invariant under the original $\mathbb{Z}_2$ symmetry of the problem, \textit{i.e.} $\theta\rightarrow \pi-\theta$. Second, and in particular, each term diverges as $\theta\rightarrow 0$. We believe both these issues originate from the perturbative setup in Brinkmann coordinates adapted to one of the shocks; thus the perturbative setup formulation does not have the $\mathbb{Z}_2$ symmetry \textit{explicitly}, and is well defined around $\theta=\pi$. The symmetry, however, must be recovered when the whole series is summed. As such, writing $\cos\theta=-\sqrt{1-\sin^2\theta}$ one must be able to expand \eqref{series2} in a $\sin^2\theta$ series (without divergent terms):
\begin{eqnarray}
\epsilon(\theta)
&\equiv&\sum_{n=0}^\infty \epsilon_n (\sin\theta)^{2n}\,.\label{angular_series}
\label{series3}
\end{eqnarray}
One can then check (using Eq.~\eqref{series2}) that
\begin{equation}
\epsilon_0=\frac{1}{8}\epsilon^{(1)}\left(\frac{\pi}{2}\right) \ , \qquad \epsilon_1=\frac{1}{32}\left[3\epsilon^{(1)}\left(\frac{\pi}{2}\right)+\epsilon^{(2)}\left(\frac{\pi}{2}\right)\right] \ , \qquad \epsilon_2=\frac{1}{128}\left[9\epsilon^{(1)}\left(\frac{\pi}{2}\right)+5\epsilon^{(2)}\left(\frac{\pi}{2}\right)+\epsilon^{(3)}\left(\frac{\pi}{2}\right)\right] \ ,
\end{equation}
and so on. Thus, a consistent truncation of the angular series~\eqref{angular_series} at some order $\mathcal{O}(n)$ requires the knowledge of $\epsilon^{(n+1)}$ and, from Section \ref{formula}, this requires knowledge of the metric perturbations up to $\mathcal{O}(n+1)$.
This observation clarifies the meaning of the perturbative expansion (\ref{eq:pertexpansion}) - it is an angular expansion off the axis of the collision - and makes explicit the correspondence between the coefficients of the two series. Then the inelasticity, in terms of the $\epsilon_n$, is
\begin{eqnarray}
\epsilon&=&\sum_{n=0}^\infty \epsilon_n \int_{-1}^1\frac{d\cos\theta}{2}(\sin\theta)^{2n}\,,\nonumber \\
&=&\sum_{n=0}^\infty \frac{2^n n!}{(2n+1)!!}\epsilon_n \\
&=&\epsilon_0+\frac{2}{3}\epsilon_1 +\dots \,. \nonumber
\label{tela}
\end{eqnarray}

Finally, an obvious corollary is that in first order perturbation theory one can only compute the expansion \eqref{series3} to zeroth order, \textit{i.e.} $\epsilon_\textrm{1st order}=\epsilon_0$. In other words, no $\theta$ dependence should be considered. This justifies the isotropy assumption used in first order perturbation theory in~\cite{Herdeiro:2011ck,Coelho:2012sya}.

\subsection{Simplification of the integrals in Fourier space}
\label{angular_factorization}
We shall now obtain simplified expressions for the integrals which provide the asymptotic metric functions that contribute to the news function (and hence to the inelasticity). By working in Fourier space with respect to the retarded time $\tau$, we shall also confirm, explicity, that the coordinate transformation \eqref{tau_transform} factorizes the angular dependence out of the integrals (asymptotically). As this section describes a long and technical computation, we have moved a considerable amount of details to Appendix \ref{app_asymF}. In this way we hope that the reader can grasp the main steps of the computation without getting distracted by the (many) details -- which are nevertheless provided in the appendix.

We start from the generic solution~\eqref{general_F}, where $F$ is now taken to represent either of the radiative metric functions $H$ or $E$ appearing in~\eqref{E}, at some order $k$ in spacetime perturbation theory.  In Appendices~\ref{a1}-\ref{a2} we simplify the (asymptotic) wave form, defined as
\begin{equation}
\dot{F}(\tau,\theta)\equiv\lim_{r\rightarrow\infty} r\rho^{\frac{D-4}{2}}\frac{d}{d\tau}F(u,v,\rho)\,.
\label{fder}
\end{equation}
In such derivation it is crucial to work in Fourier space through the definition
\begin{equation}
\hat{\dot{F}}(\omega,\theta)=\int d\tau\,\dot{F}(\tau,\theta)e^{-i\omega\tau}\,.
\label{FourierF}
\end{equation}
The simplified result is 
\begin{equation}
\hat{\dot{F}}(\omega,\theta)=-i^{\frac{D-2}{2}+m}\omega\int_0^\infty d\rho\,\rho^{\frac{D-2}{2}}J_{\frac{D-4}{2}+m}(\omega\rho\sin\theta)\hat{S}\left(\omega\frac{1+\cos\theta}{2},\omega\frac{1-\cos\theta}{2};\rho\right)\,,\label{initial_volume}
\end{equation}
where $J_\nu$ is the Bessel function of the first kind and $\hat{S}$ is the (double) Fourier transform of the source:
\begin{equation}
\hat{S}(x,y;\rho)=\frac{1}{2}\int du\int dv\, e^{-i\sqrt{2}ux}e^{-i\sqrt{2}vy}\,S(u,v,\rho)\,.\label{S_hat}
\end{equation}

We now confirm the factorization of the angular dependence proved in Section \ref{secfact}. First, let us define a new frequency space in $\Omega$ by transforming $\omega$ to
\begin{equation}
\omega\rightarrow\Omega\equiv\omega^{D-3}\left(\frac{\sin\theta}{2}\right)^{D-4}\frac{1-\cos\theta}{2}\,,\label{omega_transformation}
\end{equation}
together with a new wave form
\begin{equation}
\hat{\mathcal{F}}(\Omega,\theta)\equiv\sqrt{\omega'(\Omega)}\hat{\dot{F}}(\omega(\Omega),\theta)\,,
\label{newwaveform}
\end{equation}
such that
\begin{equation}
\int d\omega\, |\hat{\dot{F}}(\omega,\theta)|^2=\int d\Omega\, |\hat{\mathcal{F}}(\Omega,\theta)|^2\,.
\end{equation}
Applying this transformation to~\eqref{initial_volume} and after some lengthy, but straightforward, manipulations detailed in Appendix \ref{app_angular_dep} (where the CL symmetry is used) we obtain (\textit{cf.} \eqref{factorext})
\begin{equation}
\hat{\mathcal{F}}(\Omega,\theta)=\left(\frac{1}{1-\cos\theta}\right)^{\frac{3}{2}}\left(\frac{1+\cos\theta}{1-\cos\theta}\right)^{k-1}\hat{\mathcal{F}}\left(\Omega,\frac{\pi}{2}\right)\,, \label{Eq:angular_factorization}
\end{equation}
where the angular dependence is now completely factored out of the integral. Observe the $k$-dependent exponent, and hence the dependence on the order of the spacetime perturbation theory. The (new) Fourier space wave form evaluated at $\theta=\tfrac{\pi}{2}$, which we shall abbreviate to $\hat{\mathcal{F}}(\Omega)\equiv\hat{\mathcal{F}}\left(\Omega,\frac{\pi}{2}\right)$, is then given by
\begin{equation}
\hat{\mathcal{F}}(\Omega)\equiv -\sqrt{\frac{8}{D-3}}i^{\frac{D-2}{2}+m}\,\Omega^{2k-1}\int_0^\infty dR\,R^{\frac{D-2}{2}}J_{\frac{D-4}{2}+m}(2R)\hat{S}(\Omega^{-1},\Omega;R) \; . \label{Omega_form}
\end{equation}
This formulation in the new Fourier space will allow us to compute the surface terms, analytically, in Section~\ref{surface}.

Now that the angular dependence has been factored out, we can invert the transformation~\eqref{omega_transformation} at $\theta=\frac{\pi}{2}$, i.e.
\begin{equation}
\Omega\rightarrow\bar{\omega}\equiv2\,\Omega^{\frac{1}{D-3}}\,,\qquad \hat{\mathcal{F}}(\Omega)\rightarrow\sqrt{\Omega'(\bar{\omega})}\hat{\mathcal{F}}\left(\Omega(\bar{\omega})\right)\,.
\end{equation}
The net relationship between $\omega$ and $\bar{\omega}$ is
\begin{equation}
\bar{\omega}=\omega\times(1-\cos\theta)^{\frac{1}{D-3}}(\sin\theta)^{\frac{D-4}{D-3}}\,,
\end{equation}
which is equivalent, in real space, to a transformation of the time coordinate
\begin{equation}
\tau\rightarrow\bar{\tau}(\tau,\theta)=\tau\times(1-\cos\theta)^{-\frac{1}{D-3}}(\sin\theta)^{-\frac{D-4}{D-3}}\,.
\end{equation}
Apart from a $\theta$-dependent shift in $D=4$ (indeed an example of a \emph{supertranslation}~\cite{Coelho:2012sy}) $\bar{\tau}$ is exactly the same as in \eqref{tau_transform2}.

\section{Analytic evaluation of the surface terms}
\label{surface}
We are now ready to evaluate the surface terms, which exist only at first and second order in perturbation theory, and are generically given by the second term of~\eqref{general_F}.

The surface term contributions to the wave form consist of the propagation of the initial data from $u=0$ to an observation point with the Green function. In~\cite{Herdeiro:2011ck} we have shown that these boundary conditions have the general form (\textit{cf.} (3.3)-(3.6) therein; $k=1,2$ for each order)
\begin{equation}
F^{(k)}(0,v,\rho)=f(\rho)(\sqrt{2}v-\Phi(\rho))^k\theta(\sqrt{2}v-\Phi(\rho))\,.
\end{equation}
The functions $f(\rho)$ are explicitly given in~\cite{Herdeiro:2011ck} and the relevant terms will also be quoted in the next subsections; $\Phi(\rho)$ is given in \eqref{Phi}.

In Appendix \ref{app_surface2}, we show that the factored out wave form \eqref{Omega_form} simplifies to a single integral
\begin{equation}
\hat{\mathcal{F}}(\Omega)=-\sqrt{\frac{8}{D-3}}k!i^{\frac{D-2}{2}+m-k}\Omega^{k-1}\int_0^\infty dR\,R^{\frac{D-2}{2}}J_{\frac{D-4}{2}+m}(2R)f(R)e^{-i\Omega\Phi(R)}\,.
\label{app_Omega}
\end{equation}
Then, to proceed, it is convenient to Fourier transform again,
\begin{equation}
\mathcal{F}(t)=\int \dfrac{d\Omega}{2\pi}\, \hat{\mathcal{F}}(\Omega) e^{i\Omega t}\,,
\label{newft}
\end{equation}
to a time coordinate $t$ which is not proportional to $\tau$ (except in $D=4$), given the non-linear relationship between $\omega$ and $\Omega$,~\eqref{omega_transformation}. $\mathcal{F}$ is, for all practical purposes, an equivalent representation of the wave form. This transformation is not necessary to compute the inelasticity, but it is another useful representation of the wave form which can be written in closed form (\textit{cf.} \eqref{ftt}):
\begin{equation}
\mathcal{F}(t)=\sqrt{\frac{8}{D-3}}(-1)^kk!i^{\frac{D-4}{2}+m}\left[\frac{1}{\Phi'(R)}\frac{d}{dR}\right]^{k-1}\left(\Phi'(R)^{-1}R^{\frac{D-2}{2}}J_{\frac{D-4}{2}+m}(2R)f(R)\right)\,,\label{final_surface}
\end{equation}
where $R=\Phi^{-1}(t)$. To summarize, for each $E^{(k)}$, $H^{(k)}$ one must compute the corresponding $\mathcal{F}$ via \eqref{final_surface}. They will be denoted, respectively,
\[
\mathcal{F}_{E^{(k)}}(t) \ ,  \qquad \mathcal{F}_{H^{(k)}}(t) \ .
\]
Then, from  (\ref{E}) and \eqref{seriesi} we obtain an expression for $\epsilon^{(N)}(\theta)$, which, by virtue of \eqref{seriesi2} and \eqref{Eq:angular_factorization}, yields a workable formula for $\epsilon^{(N)}(\pi/2)$: the contribution to the inelasticity $\epsilon^{(N)}(\pi/2)$, at some order $N$ (which can only be $N=1,2,3$ for surface terms), is
\begin{equation}
\epsilon^{(N)}\left(\frac{\pi}{2}\right)=\frac{(D-2)(D-3)}{4} \sum_{k=1}^{N} \int dt\, \left[\mathcal{F}_{E^{(k)}}(t)+\mathcal{F}_{H^{(k)}}(t)\right]\left[\bar{\mathcal{F}}_{E^{(N+1-k)}}(t)+\bar{\mathcal{F}}_{H^{(N+1-k)}}(t)\right] \ .
\label{epsn}
\end{equation}
%
%
%
Moreover, from \eqref{final_surface}, we have, for instance, 
\begin{equation}
\int dt\, \mathcal{F}_{H^{(k)}}(t)\bar{\mathcal{F}}_{E^{(N+1-k)}}(t)=\int_0^\infty dR\,|\Phi'(R)| \mathcal{F}_{H^{(k)}}(\Phi(R)) \bar{\mathcal{F}}_{E^{(N+1-k)}}(\Phi(R))\,,\label{F_inelasticity}
\end{equation}
and similarly for the other contributions.  Thus we arrive at the striking conclusion that all surface integral contributions to the inelasticity are given by integrals over Bessel functions.


\subsection{First order contribution}
\label{first_order}
The first order estimate for the inelasticity is obtained by truncating the expansion~\eqref{tela} at leading order:
\begin{equation}
\epsilon_{\rm 1st \ order}=\epsilon_0=\frac{1}{8}\epsilon^{(1)}\left(\frac{\pi}{2}\right) .
\label{order1}
\end{equation}
To compute $\epsilon^{(1)}\left(\frac{\pi}{2}\right) $ we first observe that the trace $H^{(1)}=0$, so the initial data are specified by
\begin{equation}
E^{(1)}:\qquad k=1\,,\qquad m=2\,,\qquad f(\rho)=-\frac{\Phi'(\rho)}{\rho}=-\frac{2}{\rho^{D-2}}\,.
\end{equation}
Inserting in \eqref{final_surface} we get for the only non-trivial term
%
\begin{equation}
\mathcal{F}_{E^{(1)}}(t)=-i^{\frac{D}{2}}\sqrt{\frac{8}{D-3}} \left(R^{\frac{D-4}{2}}J_{\frac{D}{2}}(2R)\right)_{R=\Phi^{-1}(t)}\, ;
\end{equation}
using this expression in~\eqref{epsn} and making use of \eqref{F_inelasticity}, together with
%
\begin{equation}
\int_0^\infty dR\, \frac{1}{R}J_{\frac{D}{2}}(2R)^2=\frac{1}{D}\,,
\end{equation}
%
yields the result
\begin{equation}
\epsilon^{(1)}\left(\frac{\pi}{2}\right)=\frac{4(D-2)}{D}\,.
\end{equation}
Thus, from \eqref{order1}
\begin{equation}
\epsilon_{\rm 1st \ order}=\frac{1}{2}-\frac{1}{D}\,.
\end{equation}
This proves \eqref{miracle}, that was found  in~\cite{Coelho:2012sya} as an accurate numerical fit.

\subsection{Second order contributions}
\label{second_order}
At $\mathcal{O}(2)$ the initial data are specified by
\begin{eqnarray}
&E^{(2)}:&k=2\,,\qquad m=2\,,\qquad f(\rho)=-(D-4)\frac{\Phi'(\rho)^2}{4\rho^2}=-(D-4)\rho^{-2(D-2)}\\
&H^{(2)}:&k=2\,,\qquad m=0\,,\qquad f(\rho)=(D-3)\frac{\Phi'(\rho)^2}{4\rho^2}=(D-3)\rho^{-2(D-2)}\,.
\end{eqnarray}
The contribution of these terms to the inelasticity is computed in Appendix \ref{app_inelasticity}. The precise expression is of limited interest without the volume terms that also contribute at second order. Furthermore, some of the integrals do not converge for all $D\geq4$ which is potentially worrisome. To understand why, let us consider the late time behaviour of $\mathcal{F}(t)$. The far future $t\rightarrow\infty$ corresponds to $R\rightarrow0$ since
\begin{equation}
\Phi^{-1}(t)\propto\left\{
\begin{array}{ll}
 e^{-\frac{t}{2}}\ , &  D=4\  \vspace{2mm}\\
\displaystyle{ t^{-\frac{1}{D-4}}}\ , & D>4\ \label{Phi2}
\end{array} \right. \ .
\end{equation} 
The asymptotic behaviour of the wave forms is
\begin{equation}
\mathcal{F}_{E^{(1)}}(t)\simeq R^{D-2}\,,\qquad \mathcal{F}_{E^{(2)}}(t)\simeq R^{D-2}\,,\qquad \mathcal{F}_{H^{(2)}}(t)\simeq R^{D-6}\,,
\end{equation}
which means that $\mathcal{F}_{H^{(2)}}$ grows exponentially with $e^t$ in $D=4$. Indeed, only for $D>8$ does it decay faster than $t^{-1}$ which is the condition for integrability. This is in agreement with the findings of D'Eath and Payne~\cite{DEath:1992qu} in $D=4$: both the surface and volume terms have non-integrable, exponentially growing tails at late times, but their sum is well behaved and integrable. In Appendix \ref{app_tails} we trace the origin of these tails to the Green's function and show that they are generic. In particular, we confirm that the volume terms of $E^{(2)}$ and $H^{(2)}$ have the same behaviour as their surface counterparts above. Therefore we expect (though without proof) a similar cancellation in $D>4$.

In Table~\ref{TabSurface}, we summarise all the surface integral contributions which arise from first and  second order spacetime perturbations.
The angular dependence was suppressed since it has already been factorized and identified. We have checked a very good agreement between all these results -- in the domain of $D$ where the integrals are convergent, \textit{i.e.} the tails die off fast enough -- and the numerical code we have developed previously for the surface integrals, which consisted of a completely different approach~\cite{Herdeiro:2011ck,Coelho:2012sya,Coelho:2012sy}. They agree with a relative error of less than $10^{-4}$.

\begin{table}[h!]
\begin{center}
\begin{tabular}{||c|c|c||}
\hline
 $N$ & Term & contribution to $\epsilon^{(N)}(\tfrac{\pi}{2})$ \\
\hline
\hline
1 &  $\mathcal{F}_{E^{(1)}}\mathcal{F}_{E^{(1)}}$& $8\left(\frac{1}{2}-\frac{1}{D}\right)\phantom{\dfrac{D}{D}}$\\
\hline
2 &$2\mathcal{F}_{E^{(1)}}\mathcal{F}_{E^{(2)}}$ & $-32\left(\frac{1}{2}-\frac{1}{D}\right)\tfrac{D-4}{D+2}\phantom{\dfrac{D}{D}}$ \\
& $2\mathcal{F}_{E^{(1)}}\mathcal{F}_{H^{(2)}}$& $-32\left(\frac{1}{2}-\frac{1}{D}\right)\tfrac{D-3}{D-4}\phantom{\dfrac{D}{D}}$\\
\hline
3    & $\mathcal{F}_{E^{(2)}}\mathcal{F}_{E^{(2)}}$ & $64\left(\frac{1}{2}-\frac{1}{D}\right)\tfrac{(D-4)^2}{(D+2)(D+4)}\phantom{\dfrac{D}{D}}$\\
     & 2$\mathcal{F}_{E^{(2)}}\mathcal{F}_{H^{(2)}}$ & $64\left(\frac{1}{2}-\frac{1}{D}\right)\tfrac{(D-3)}{(D+2)}\phantom{\dfrac{D}{D}}$\\
    & $\mathcal{F}_{H^{(2)}}\mathcal{F}_{H^{(2)}}$ & $64\left(\frac{1}{2}-\frac{1}{D}\right)\tfrac{(D-3)^2}{(D-4)(D-8)}\phantom{\dfrac{D}{D}}$\\
\hline
\end{tabular}\vspace{2mm}\\
\end{center}
\caption{\label{TabSurface} Surface integral contributions in~\eqref{epsn} to the angular series used to compute the inelasticity.}
\end{table}

\section{Conclusions}
\label{conclusions}
In this paper, we have considered the collision of two $D$-dimensional Aichelburg-Sexl shock waves, following our previous work~~\cite{Herdeiro:2011ck,Coelho:2012sya,Coelho:2012sy} (see also the summaries in~\cite{Coelho:2013zs,Coelho:2014qla,Coelho:2014cna} and the lecture notes in~\cite{Sampaio:2013faa}), extending the perturbative framework of D'Eath and Payne to higher dimensions. 

Our first result was to demonstrate that the CL symmetry, which is a conformal isometry  of the geometry of two colliding shock waves -- order by order in perturbation theory --  implies that the angular dependence factorizes at null infinity. Note that this is a purely kinematical result, not depending on the equations of motion. Moreover, that dependence has a simple form at all orders, both for surface and volume terms. As a byproduct, we clarified the meaning of perturbation theory -- it can be seen as an angular expansion around the symmetry axis. Finally, and perhaps our main result here, we computed the inelasticity in first order perturbation theory by purely analytic means, thus demonstrating \eqref{miracle}, which was derived in \cite{Coelho:2012sya} as a numerical fit. 

Our treatment also demonstrated that both surface and volume terms at second order may contain non-integrable tails (unless $D>8$). There is a realistic hope that they will cancel one another in higher $D$ as they do in $D=4$~\cite{DEath:1992qu}. To obtain the inelasticity in second order perturbation theory for arbitrary $D$ remains the outstanding challenge in this research program.

\vspace{0.5cm} 
\noindent
{\bf\large Acknowledgements}\\ 
F.C. and M.S. are supported by the FCT grants SFRH/BD/60272/2009 and SFRH/BPD/69971/2010. C.H gratefully acknowledges support from the FCT-IF programme.  The work in this paper is also supported by the grants PTDC/FIS/116625/2010 and  NRHEP--295189-FP7-PEOPLE-2011-IRSES.

 \newpage 
 \appendix

\section{The asymptotic wave form and its Fourier transform}
\label{app_asymF}

\subsection{The asymptotic wave form}
\label{a1}

The Green's function for the d'Alembertian operator $\Box$ in $D$ dimensions is given by (see Appendix A in~\cite{Herdeiro:2011ck})
\begin{equation}
G(x^\mu)=-\frac{1}{2}\frac{1}{\pi^\frac{D-2}{2}}\delta^{\left(\frac{D-4}{2}\right)}(\chi),\qquad \chi\equiv \eta_{\mu\nu}x^\mu x^\nu\,, \label{GreenF}
\end{equation}
Inserting this in the general solution (\ref{general_F}) and integrating over the angles $\phi_i$ on the transverse plane, we get
\begin{equation}
F(u,v,\rho)=-\frac{\Omega_{D-4}}{2\pi^{\frac{D-2}{2}}}\int du'\int dv'\int d\rho'\,\rho'^{D-3}S(u',v',\rho')\int_{-1}^1dx\,(1-x^2)^{\frac{D-5}{2}+m}\Lambda_m(x)\delta^{\left(\frac{D-4}{2}\right)}(\chi)\,.
\end{equation}
Here $\Omega_n$ is the volume of the $n$-sphere
\begin{equation}
\Omega_n=\frac{2\pi^{\frac{n+1}{2}}}{\Gamma\left(\frac{n+1}{2}\right)}\,,
\end{equation}
and we recall that $\Lambda_m$ was defined in \eqref{lm}.

Using the scaling properties of the delta function and defining $x'$ through
\begin{eqnarray}
\chi&=&2(u-u')(v-v')-(\rho^2+\rho'^2-2\rho\rho' x)\nonumber \\
&\equiv&2\rho\rho'(x-x')\,,
\end{eqnarray}
this becomes
\begin{equation}
F(u,v,\rho)=-\frac{1}{2\rho^{\frac{D-2}{2}}}\int du'\int dv'\int d\rho'\,\rho'^{\frac{D-4}{2}}S(u',v',\rho')I_m^{D,0}(x')\,,
\end{equation}
where
\begin{equation}
I_m^{D,N}(z)\equiv\frac{\Omega_{D-4}}{(2\pi)^{\frac{D-2}{2}}}\int_{-1}^1dx\,\Lambda_m(x)(1-x^2)^{\frac{D-4}{2}-\frac{1}{2}}\delta^{\left(\frac{D-4}{2}-N\right)}(x-z)\,.\label{def_I}
\end{equation}

We are actually interested in the asymptotic form of the time derivative of $F$, \eqref{fder}. To compute it
let $\bar{x}$ be the asymptotic form of $x'$,\footnote{Note that $\lim_{r\rightarrow\infty}\partial_\tau x'=\partial_\tau\lim_{r\rightarrow\infty}x'\,.
$}
\begin{equation}
\lim_{r\rightarrow\infty}x'=\frac{1}{\rho'\sin\theta}\left(-\tau+\sqrt{2}u'\frac{1+\cos\theta}{2}+\sqrt{2}v'\frac{1-\cos\theta}{2}\right)\equiv \bar{x}\,.\label{x_bar}
\end{equation}
Noting that $\left(I_m^{D,N}\right)'=-I_m^{D,N-1}$, we get
\begin{equation}
\dot{F}(\tau,\theta)=\frac{1}{2\sin\theta}\int du'\int dv'\int d\rho'\,\rho'^{\frac{D-4}{2}}S(u',v',\rho')I_m^{D,-1}(\bar{x})\frac{d \bar{x}}{d\tau}\,.
\label{fder2}
\end{equation}


\subsection{Fourier transform}
\label{a2}

Let the Fourier transform of $\dot{F}(\tau,\theta)$ with respect to the retarded time $\tau$ be defined by \eqref{FourierF}.
By inverting the function $\bar{x}(\tau)$ to $\tau(\bar{x})$, we can switch the integration variable,
\begin{equation}
\int_{-\infty}^{+\infty}d\tau\,\frac{d \bar{x}}{d\tau}(\tau)I_m^{D,-1}(\bar{x}(\tau))e^{-i\omega\tau}=\int_{+\infty}^{-\infty}d\bar{x} I_m^{D,-1}(\bar{x})e^{-i\omega\tau(\bar{x})}=-\int dx\, I_m^{D,-1}(x)e^{-i\omega\tau(x)}\,.\label{x_int}
\end{equation}
The $x$ integration in (\ref{x_int}) is essentially
\begin{equation}
\int dx\, I_m^{D,-1}(x)e^{izx}\,,\qquad z=\omega\rho'\sin\theta\,.
\end{equation}
From the definition (\ref{def_I}), this is
\begin{eqnarray}
&&\frac{\Omega_{D-4}}{(2\pi)^{\frac{D-2}{2}}}\int dy\,e^{iyz}\int_{-1}^{1} dx\,\Lambda_m(x)(1-x^2)^{\frac{D-4}{2}-\frac{1}{2}}\delta^{\left(\frac{D-2}{2}\right)}(x-y)\\
&=&\frac{\Omega_{D-4}}{(2\pi)^{\frac{D-2}{2}}}\int dy\,e^{iyz}\int_{-1}^{1} dx\,\lambda_m\partial^{(m)}(1-x^2)^{\frac{D-4}{2}+m-\frac{1}{2}}\delta^{\left(\frac{D-2}{2}\right)}(x-y)\\
&=&\frac{\Omega_{D-4}}{(2\pi)^{\frac{D-2}{2}}}(iz)^{\frac{D-2}{2}}\lambda_m\int_{-1}^{1} dx\,\partial^{(m)}(1-x^2)^{\frac{D-4}{2}+m-\frac{1}{2}}e^{ixz}\\
&=&\frac{\Omega_{D-4}}{(2\pi)^{\frac{D-2}{2}}}(iz)^{\frac{D-2}{2}+m}(-1)^m\lambda_m\int_{-1}^{1} dx\,(1-x^2)^{\frac{D-4}{2}+m-\frac{1}{2}}e^{ixz}\,,\label{poisson}
\end{eqnarray}
where
\begin{equation}
\lambda_0=1\,,\qquad \lambda_1=-(D-3)^{-1}\,,\qquad \lambda_2=(D-3)^{-1}(D-1)^{-1}\,.
\end{equation}
One recognizes in (\ref{poisson}) the Poisson integral representation of the Bessel function of the first kind,
\begin{equation}
J_\nu(z)=\frac{1}{\sqrt{\pi}\Gamma\left(\nu+\frac{1}{2}\right)}\left(\frac{z}{2}\right)^\nu\int_{-1}^1 ds\, (1-s^2)^{\nu-\frac{1}{2}}e^{isz}\,.
\end{equation}
We get
\begin{equation}
\int dx\, I_m^{D,-1}(x)e^{izx}=i^{\frac{D-2}{2}+m}zJ_{\frac{D-4}{2}+m}(z)\,.
\end{equation}
The final expression is thus \eqref{initial_volume}.
The $\rho$ integral therein can actually be cast as a $(D-2)$-dimensional Hankel transform, with argument $\omega\sin\theta$. This is not surprising since the Hankel transform arises whenever a Fourier transform is made of a function with spherical symmetry.

\subsection{Factorization of the angular dependence}
\label{app_angular_dep}
Starting with (\ref{initial_volume}), we make the following transformations
\begin{equation}
\rho\rightarrow R\equiv\frac{1}{2}\omega\rho\sin\theta\,,
\end{equation}
and \eqref{omega_transformation}. Moreover we define a new transformed wave form \eqref{newwaveform}.  
We get
\begin{eqnarray}
\hat{\mathcal{F}}(\Omega,\theta)&=&-\frac{i^{\frac{D-2}{2}+m}}{\sqrt{D-3}}\left(\frac{2}{\sin\theta}\right)^2\left(\frac{1-\cos\theta}{2}\right)^{\frac{1}{2}}\times\Omega^{-1}\nonumber \\
&\times&\int_0^\infty dR\,R^{\frac{D-2}{2}}J_{\frac{D-4}{2}+m}(2R)\hat{S}\left(\omega\frac{1+\cos\theta}{2},\omega\frac{1-\cos\theta}{2};\frac{2R}{\omega\sin\theta}\right)_{\omega=\omega(\Omega)}\,.
\end{eqnarray}
Now the CL symmetry comes in handy. From Section \ref{CL}, we know that if $F$ is $\mathcal{O}(k)$,\footnote{Note that here $F$ is either $E$ or $H$ so $N_u=N_v=0$ in (\ref{def_Sigma}).}
\begin{equation}
S(u,v,\rho)=\rho^{-2k(D-3)-2}\,\Sigma(p,q)\,,
\end{equation}
where
\begin{equation}
p=\left(\sqrt{2}v-\Phi(\rho)\right)\rho^{D-4}\,,\qquad q=\sqrt{2}u\rho^{-(D-2)}\,.
\end{equation}
Therefore, and using $2dudv=\rho^2dpdq$,
\begin{equation}
\hat{S}(x,y;\rho)=\frac{1}{4}\rho^{-2k(D-3)}e^{-iy\Phi(\rho)}\int dp\int dq\,\Sigma(p,q)\,e^{-ixq\rho^{D-2}}\,e^{-iyp\rho^{-(D-4)}}\,.
\end{equation}
Now we have
\begin{eqnarray}
\rho^{-2k(D-3)}&=&(\omega\sin\theta)^{2k(D-3)}(2R)^{-2k(D-3)}\,,\\
y\Phi(\rho)&=&\Omega\Phi(R)\,,\\
x\rho^{D-2}&=&\Omega^{-1}R^{D-2}\,,\\
y\rho^{-(D-4)}&=&\Omega R^{-(D-4)}\,.
\end{eqnarray}
Thus,
\begin{eqnarray}
&&\hat{S}\left(\omega\frac{1+\cos\theta}{2},\omega\frac{1-\cos\theta}{2};\frac{2R}{\omega\sin\theta}\right)=\nonumber\\
&=&\frac{1}{4}\frac{(\omega\sin\theta)^{2k(D-3)}}{(2R)^{2k(D-3)}}\int dp\int dq\,\Sigma(p,q)\,e^{-iq\frac{R^{D-2}}{\Omega}}\,e^{-i\Omega\left(\Phi(R)+\frac{p}{R^{D-4}}\right)}\,,
\end{eqnarray}
and finally,
\begin{eqnarray}
\hat{\mathcal{F}}(\Omega,\theta)&=&-\frac{i^{\frac{D-2}{2}+m}}{\sqrt{D-3}}\times\left(\frac{1+\cos\theta}{1-\cos\theta}\right)^{k-1}\left(\frac{2}{1-\cos\theta}\right)^{\frac{3}{2}}\times\Omega^{2k-1}\times\nonumber\\
&\times&\frac{1}{4}\int_0^\infty dR\,R^{\frac{D-2}{2}-2k(D-3)}J_{\frac{D-4}{2}+m}(2R)\times\nonumber\\
&\times&\int dp\int dq\,\Sigma(p,q)\,e^{-iq\frac{R^{D-2}}{\Omega}-i\Omega\left(\Phi(R)+\frac{p}{R^{D-4}}\right)}\,\,.\label{app_transformed_F}
\end{eqnarray}
At this point, the $p,q$ coordinates are not necessarily advantageous since the $R$ integration cannot be done as it is. So we might as well go back to
\begin{equation}
\sqrt{2}u=qR^{D-2}\,,\qquad \sqrt{2}v=pR^{-(D-4)}+\Phi(R)\,,
\end{equation}
and rewrite (\ref{app_transformed_F}) as
\begin{eqnarray}
\hat{\mathcal{F}}(\Omega,\theta)&=&-\frac{i^{\frac{D-2}{2}+m}}{\sqrt{D-3}}\left(\frac{2}{1-\cos\theta}\right)^{\frac{3}{2}}\left(\frac{1+\cos\theta}{1-\cos\theta}\right)^{k-1}\times \nonumber \\
&\times&\Omega^{2k-1}\int_0^\infty dR\,R^{\frac{D-2}{2}}J_{\frac{D-4}{2}+m}(2R)\hat{S}(\Omega^{-1},\Omega;R)\,.
\label{factorext}
\end{eqnarray}

\section{Computation of surface terms}
\label{app_surface}

\subsection{Introducing the source}
\label{app_surface2}

The source for the surface terms is
\begin{equation}
S^{(k)}(u,v,\rho)=2\delta(u)\partial_vF^{(k)}(0,v,\rho)\,,\qquad F^{(k)}(0,v,\rho)=f(\rho)(\sqrt{2}v-\Phi(\rho))^k\theta(\sqrt{2}v-\Phi(\rho))\,,
\end{equation}
where $k=\{1,2\}$ and $f(\rho)\propto\rho^{-k(D-2)}$ for $E$ and $H$. Then, using the definition (\ref{S_hat}), we get
\begin{equation}
\hat{S}(\Omega^{-1},\Omega;R)=\frac{k!}{(i\Omega)^k}f(R)e^{-i\Omega\Phi(R)}\,.
\end{equation}
Inserting in (\ref{Omega_form}) we find \eqref{app_Omega}.
As explained in the main text we then Fourier transform again, \textit{cf.} \eqref{newft}, back to a time coordinate $t$.
This new wave form is not exactly the original $\dot{F}(\tau,\theta)$. To perform this integral, we change the integration variable in \eqref{app_Omega} from $R$ to $T=\Phi^{-1}(R)$, i.e.
\begin{equation}
\int_0^\infty dR\simeq \int_{\Phi(0)}^{\Phi(\infty)} dT\,\Phi'(R)^{-1}\,,\qquad R=\Phi^{-1}(T)\,.
\end{equation}
Then the $\Omega$ integral is
\begin{equation}
\frac{1}{2\pi}\int d\Omega\, \Omega^{k-1} e^{-i\Omega(T-t)}=i^{k-1}\delta^{(k-1)}(T-t)\,,
\end{equation}
yielding
\begin{eqnarray}
\mathcal{F}(t)&=&-\sqrt{\frac{8}{D-3}}k!i^{\frac{D-4}{2}+m}\int_{\Phi(0)}^{\Phi(\infty)} dT\,\Phi'(R)^{-1}R^{\frac{D-2}{2}}J_{\frac{D-4}{2}+m}(2R)f(R)\delta^{(k-1)}(T-t)\,,\\
&=&\sqrt{\frac{8}{D-3}}(-1)^kk!i^{\frac{D-4}{2}+m}\left[\frac{1}{\Phi'(R)}\frac{d}{dR}\right]^{k-1}\left(\Phi'(R)^{-1}R^{\frac{D-2}{2}}J_{\frac{D-4}{2}+m}(2R)f(R)\right)\,, \label{ftt}
\end{eqnarray}
where $R=\Phi^{-1}(t)$.

\subsection{Surface terms contribution to the inelasticity}
\label{app_inelasticity}
When considering the integrated product of $(\dot{E}+\dot{H})(\bar{\dot{E}}+\bar{\dot{H}})$ (the bar denotes complex conjugation), we have at each order
\begin{eqnarray}
\mathcal{O}(2): &\qquad& \dot{E}^{(1)}\bar{\dot{E}}^{(1)}\,,\\
\mathcal{O}(3): &\qquad& 2 Re\left[\dot{E}^{(1)}\bar{\dot{E}}^{(2)}\right]+2 Re\left[\dot{E}^{(1)}\bar{\dot{H}}^{(2)}\right]\,,\\
\mathcal{O}(4): &\qquad& \dot{E}^{(2)}\bar{\dot{E}}^{(2)}+\dot{H}^{(2)}\bar{\dot{H}}^{(2)}+2Re\left[\dot{E}^{(2)}\bar{\dot{H}}^{(2)}\right]\,.
\end{eqnarray}

Below we list the integrals involving Bessel functions that arise for each of the above terms.
\begin{eqnarray}
\dot{E}^{(1)}\bar{\dot{E}}^{(1)}: &\quad& \int_0^\infty dR\, \frac{1}{R}J_{\frac{D}{2}}(2R)^2=\frac{1}{D}\,,\quad D>0\,,\nonumber \\
2 Re\left[\dot{E}^{(1)}\bar{\dot{E}}^{(2)}\right]: &\quad& \int_0^\infty dR\, \frac{1}{R^2}J_{\frac{D}{2}}(2R)J_{\frac{D}{2}+1}(2R)=\frac{2}{D}\frac{1}{D+2}\,,\quad D>0\,,\nonumber \\
2 Re\left[\dot{E}^{(1)}\bar{\dot{H}}^{(2)}\right]: &\quad& \int_0^\infty dR\, \frac{1}{R^3}J_{\frac{D}{2}}(2R)\left(J_{\frac{D-4}{2}}(2R)+R\, J_{\frac{D-2}{2}}(2R)\right)=\frac{2}{D}\frac{1}{D-4}\,,\quad D>4\,,\nonumber \\
\dot{E}^{(2)}\bar{\dot{E}}^{(2)}: &\quad& \int_0^\infty dR\, \frac{1}{R^3}J_{\frac{D}{2}+1}(2R)^2=\frac{8}{D}\frac{1}{(D+2)(D+4)}\,,\quad D>0\,,\nonumber \\
\dot{H}^{(2)}\bar{\dot{H}}^{(2)}: &\quad& \int_0^\infty dR\,\frac{1}{R^4}J_{\frac{D}{2}+1}(2R)\left(J_{\frac{D-4}{2}}(2R)+R\,J_{\frac{D-2}{2}}(2R)\right)=\frac{4}{D}\frac{1}{(D-4)(D+2)}\,,\quad D>4\,,\nonumber \\
2Re\left[\dot{E}^{(2)}\bar{\dot{H}}^{(2)}\right]: &\quad& \int_0^\infty dR\, \frac{1}{R^5}\left(J_{\frac{D-4}{2}}(2R)+R\,J_{\frac{D-2}{2}}(2R)\right)^2=\frac{8}{D}\frac{1}{(D-4)(D-8)}\,,\quad D>8\,. \nonumber
\end{eqnarray}

Their contributions to $\epsilon\left(\tfrac{\pi}{2}\right)$ can be found in Table~\ref{TabSurface}.

\subsection{The late time tails}
\label{app_tails}
Following the strategy of Section \ref{surface} for the surface terms, we perform an inverse Fourier transform from $\Omega$ to a new time coordinate $t$. The relevant integral is
\begin{equation}
\frac{1}{2\pi}\int d\Omega\, \Omega^{2k-1}e^{-i\Omega P-iQ\Omega^{-1}}=(-i\partial_t)^{2k-1}\frac{1}{2\pi}\int d\Omega\, e^{-i\Omega P-iQ\Omega^{-1}}\,,
\end{equation}
where
\begin{equation}
P=\Phi(R)+pR^{-(D-4)}-t\,,\qquad Q=qR^{D-2}\,.
\end{equation}
This integral simplifies to
\begin{equation}
i^{-2k-1}\partial_t^{(2k)}\,\theta(X)J_0(2\sqrt{X})\,,\qquad X\equiv PQ\, .
\end{equation}
Therefore, we get
\begin{equation}
\mathcal{F}(t,\theta)\propto \int dp\int dq\,\Sigma(p,q)\, G(p,q;t)\,,
\end{equation}
where $G(p,q;t)$ is essentially the asymptotic (i.e. for an observer at null infinity) Green's function in $p,q$ coordinates and for the new time $t$:
\begin{equation}
G(p,q;t)=\left(\frac{d}{dt}\right)^{2k}\int_0^\infty dR\,R^{\frac{D-2}{2}-2k(D-3)}J_{\frac{D-4}{2}+m}(2R)\theta(X)J_0(2\sqrt{X})\,,\label{G_pqt}
\end{equation}
where
\begin{eqnarray}
X\equiv PQ=qR^{D-2}\left(\left(p+\frac{2}{D-4}\right)R^{-(D-4)}-t\right)\,.
\end{eqnarray}
The action of the $2k$ derivatives distributes as
\begin{eqnarray}
&&\left(qR^{D-2}\right)^{2k}\sum_{j=0}^{2k}{2k\choose j}\frac{(-1)^j}{j!X^{\frac{j}{2}}}\delta^{(2k-j-1)}(X)J_j\left(2\sqrt{X}\right)\\
&=&\left[\sum_{j=0}^{2k-1}{2k\choose j}\frac{(-qR^{D-2})^j}{j!}\delta^{(2k-j-1)}(P)\right]+\frac{\left(qR^{D-2}\right)^{2k}}{(2k)!X^k}\theta(X)J_{2k}\left(2\sqrt{X}\right)\,.
\end{eqnarray}
The $R$ integral in (\ref{G_pqt}) can be readily done for all but the very last term. So the contribution of the series inside the square brackets for $G(p,q;t)$ is
\begin{equation}
\sum_{j=0}^{2k-1}{2k\choose j}\frac{q^j}{\left(\frac{D-4}{2}p+1\right)^{2k-j}}\left(\frac{1}{\Phi'(T)}\frac{d}{dR}\right)^{2k-j-1}\left(\frac{R^{\frac{D-2}{2}-2k(D-3)+j(D-2)}}{\Phi'(R)}J_{\frac{D-4}{2}+m}(2R)\right)\,,\label{series_G_pqt}
\end{equation}
where
\begin{equation}
R=\left\{
\begin{array}{ll}
\Phi^{-1}(t-p)\ , &  D=4\  \vspace{2mm}\\
\displaystyle{\Phi^{-1}\left(t\left(\frac{D-4}{2}p+1\right)^{-1}\right)}\ , & D>4\,
\end{array} \right. \ .
\end{equation} 
Late times $t\gg 1$ correspond to $R\ll 1$. Thus we can study the late-time tails by making a series expansion of (\ref{series_G_pqt}) for $R\simeq0$. The most relevant term is $j=0$, since it has the highest powers of $R$ (indeed, for $k=1$ this is the only surviving term because of the $\delta(q)$ in $\Sigma(p,q)$). The wave form will only be square-integrable if its tail decays at least with $t^{-1}$. We recover the results of section \ref{second_order} : $E^{(1)}$ and $E^{(2)}$ are square integrable for all $D\geq4$, but $H^{(2)}$ is not: it grows exponentially in $D=4$ and with $t^{-\alpha}$ for higher $D$. The condition $\alpha\geq1$ is only true for $D>8$. 


\newpage

 \bibliographystyle{h-physrev4}
\bibliography{hbhs}

\end{document}